\documentstyle[psfig,twocolumn,prl,aps,epsf]{revtex}

\def\be{\begin{equation}}
\def\ee{\end{equation}}
\def\bea{\begin{eqnarray}}
\def\eea{\end{eqnarray}}

\begin{document}

\title{Moments of the Proton $F_2$ Structure Function at Low $Q^2$}
\author{C.S.~Armstrong,$^{1}$ R.~Ent,$^{1}$ C.E.~Keppel,$^{1,2}$
S.~Liuti,$^{3,4}$ G.~Niculescu,$^{5}$ I.~Niculescu$^{1,6}$}

\address{
$^{1}$ Thomas Jefferson National Accelerator Facility, Newport News, Virginia 23606
$^{2}$ Hampton University, Hampton, Virginia 23668
$^{3}$ University of Virginia, Charlottesville, Virginia 22901
$^{4}$ INFN Sezione di Roma Tre, Roma, Italy
$^{5}$ Ohio University, Athens, Ohio 45071
$^{6}$ The George Washington University, Washington, D.C. 20052}
\newpage
\date{\today}
\maketitle

\begin{abstract}
We investigate the $Q^2$ dependence of inclusive electron-proton scattering
$F_2$ structure function data in both the nucleon resonance region and the
deep inelastic region, at momentum transfers below 5 (GeV/c)$^2$.
From these data we construct moments of $F_2$, down
to momentum transfers of $Q^2 \approx$ 0.1 (GeV/c)$^2$.
The second moment is only slowly varying with $Q^2$
down to $Q^2 \approx$ 1 (GeV/c)$^2$, which is
a reflection of duality. Below $Q^2$ of 1 (GeV/c)$^2$, the $Q^2$
dependence of the moments is predominantly governed by the elastic
contribution, whereas the inelastic channels still seem governed by
local duality.
\end{abstract}



\section{Introduction}

Perturbative QCD (pQCD) describes the scale dependence
of a wide number of hard processes, up to Next to Leading Order (NLO),
and in a few cases up to Next to Next to Leading Order (NNLO) \cite{review}. 
However, direct comparisons of calculations with experimental cross sections
are often affected by the uncertainty coming from non-perturbative 
contributions, that are less well understood,
although exceptional cases exist where the non-perturbative corrections can be
deduced
from the large distance behavior of the perturbative series \cite{BenekePR}.
To this date, the mechanisms by which 
a pQCD description of deep inelastic observables starts failing,
giving way to non-perturbative (NP) behavior, are still
largely undetermined. New studies of kinematic regions,
and possible observables expected to be most sensitive to this transition,
are currently being pursued.
  
This paper is dedicated to interpreting a somewhat surprising behavior
of the nucleon structure function $F_2$ in the nucleon resonance region,
as recently measured at Jefferson Lab and reported in \cite{ioana1,ioana2},
for four-momentum transfers squared 0.45 $\leq Q^2 \leq$ 3.3 (GeV/c)$^2$.
This behavior is consistent with older, lower precision, 
SLAC data, down to $Q^2$ = 0.07 (GeV/c)$^2$ \cite{stein}.
At the lower values of $Q^2$, the resonance region, typically defined to
comprise the region in invariant mass
$M^2 < W^2$ (= $M^2$ + $Q^2$(1/$x$ - 1)) $<$ 4 GeV$^2$, with $M$ the
proton mass and $x$ the Bjorken scaling variable, is at low values
of $x$, $x \approx 0.1$. Thus, the resonances do not characterize the large $x$
behavior that was extensively studied before, in e.g. Refs. 
\cite{dukerob,abbbar,ji,twistpaper}.
It was, however, found that the resonance region seems to average to a
single curve even at these very low values of $Q^2$. This scaling curve
mimics a valence-like quark distribution, the new data of \cite{ioana1}
adding new information for $x \leq$ 0.25.

It is the aim of this paper to present a low-$Q^2$ moment analysis of the
inclusive electron-proton scattering $F_2$ structure function data, based upon
the new precise resonance region and existing deep inelastic region data,
and to interpret these moments in the context of the surprising resonance
region behavior of \cite{ioana1}. Such an analysis
would extend previous
pQCD analyses of the type performed in \cite{dgp,dukerob,abbbar,ji,twistpaper},
to the lowest values of $Q^2$ and $x$ of \cite{ioana1}.
In particular, as $Q^2$ decreases, we expect NP effects to
dominate the cross section, reducing the agreement with pQCD predictions. 
Such an analysis may be complementary to the ones addressing another set of
low-$Q^2$ measurements of $F_2$ obtained recently at HERA \cite{zeus00},
which however belong to a completely different kinematical region,
characterized by very low $x$ (down to $x \approx 10^{-6}$)
and, accordingly, to very large $W^2$.  
In \cite{zeus00} it has been found that the structure function $F_2$ 
can be described by pQCD down to $Q^2 \approx 1 $ (GeV/c)$^2$,
provided a gluon distribution vanishing  at low $x$ (also referred to in the
literature as ``valence-like'' \cite{zeus00}) and a non vanishing but small
sea distribution, are adopted.    
However, in Ref.\cite{zeus00} it was also found that, at even lower values
of $Q^2$, $0.045 \leq Q^2 \leq 0.65$ (GeV/c)$^2$, and in the same very low $x$
region, the pQCD description breaks down giving way to different types 
({\it e.g.} Regge type) of descriptions.

The fact that pQCD can be extended to very low 
$Q^2$ ($Q^2 \approx $ 1 (GeV/c)$^2$) was unexpected and supports the idea  
that the QCD coupling constant might have a smoother
behavior at low $Q^2$ than predicted by asymptotic freedom expressions.
Or, in other words, the confinement mechanism might manifest
itself in a softer way \cite{Dok,hoyer,MatSte}.
It is also unexpected from the point of view of
a pQCD based analysis that the resonance data, lying on the 
low side of the invariant mass spectrum, would 
follow a curve that is not too far from the DIS valence quark curve
down to very low $Q^2$. Is this still a signature of pQCD?     

In Section II we remind the reader of the Bloom-Gilman duality phenomenon.
In Section III we review the world
inclusive (e,e$^\prime$) data at low $Q^2$, both in the low and
in the large $W^2$ regions. In Section IV, we evaluate the moments of the
proton structure function in the low $Q^2$ region, including the elastic
contribution, and we extend the pQCD based analysis of \cite{twistpaper} to 
this region. In Section V, we discuss the role played by the elastic
contribution in rendering the $Q^2$ dependence of the moments.
In Section VI, we draw our conclusions.

\section{Bloom-Gilman Duality}

Three decades ago, Bloom and Gilman observed a fascinating correspondence
between the resonance electroproduction and deep inelastic 
kinematic regimes of inclusive electron-nucleon scattering \cite{bg}.
Specifically, it was observed that the resonance strength could be
related to the deep inelastic strength via a scaling variable which
allowed for a comparison of the lower $W^2$ and $Q^2$ resonance region data
to the higher $W^2$
and $Q^2$ deep inelastic data. It was observed that the deep inelastic
data are apparently equivalent to an average of the resonance
region data. Furthermore, this behavior was observed over a range in $Q^2$ and
$W^2$, and it was found that, with changing $Q^2$, the resonances move along,
but always average to, the smooth scaling curve typically associated with
deep inelastic scattering (DIS). This behavior clearly hinted at a common origin
for resonance (hadron) electroproduction and deep inelastic
(partonic) scattering, termed parton-hadron, or Bloom-Gilman, duality.

A global kind of parton-hadron
duality is well established: low-energy resonance production
can be shown to be related to the high-energy behavior of hadron-hadron
scattering \cite{olddual,harari,regge};
the familiar ratio of $e^+e^- \rightarrow$ hadrons
over $e^+e^- \rightarrow$ muons uses duality to relate the hadron production
to the sum of the squared charges of the quarks: here duality is guaranteed
by unitarity (in this, one could argue that the $\rho$ production channel
exhibits local duality, in that its area averages to about the same global
value) \cite{close};
in pQCD the high-momentum transfer
behavior of nucleon resonances can be related to the high-energy transfer
behavior of DIS \cite{close,roberts}.
However, it is not clear why duality should also work
in a localized region, and even at relatively low momentum transfers.

Inclusive deep inelastic
scattering on nucleons is a firmly-established tool for the investigation of
the quark-parton model. At large enough values of
$W^2$ and $Q^2$, with $W^2 >> Q^2$, a precise
description of the $Q^2$ behavior of the nucleon
structure function $F_2 = \nu W_2$ can be given 
in terms of a perturbative series in $\alpha_S(Q^2)$, up to
NLO \cite{alt,bur}. Such $Q^2$ behavior
becomes especially transparent in comparing the high $Q^2$ ($>$ 10 (GeV/c)$^2$)
moments of $F_2$ with pQCD predictions \cite{roberts,dukerob}.
 
An analysis of the resonance region at smaller $W^2$ and $Q^2$
in terms of QCD was first presented in Refs. \cite{dgp}, where Bloom and
Gilman's duality was re-interpreted: The integrals of the structure function,
performed in \cite{bg} over the energy transfer $\nu$, 
were translated into integrals over the variable $x$ (or Nachtmann
$\xi = 2x/(1+\root \of {1+4M^2x^2/Q^2}$ \cite{nacht}, in order to account for
finite target mass effects).  
Bloom-Gilman duality was translated into a correspondence 
between the $n = $2 moment of the $F_2$ structure
function in the low $Q^2$ region, characterized by resonances, and in 
the high $Q^2$ scaling region, respectively.
The fall of the resonances along a smooth scaling curve with increasing $Q^2$
was to be attributed \cite{dgp}
to the fact that there exist only small changes in the low $n$ 
moments of the $F_2$ structure function due to power corrections 
in addition to the predicted perturbative ones.
The appearence of power corrections is interpreted as a signal of
deviations of the inclusive cross section from perturbative
predictions, which one can envisage as due to the 
increasing importance of interactions between the 
quark struck in the electron-nucleon hard scattering process 
and the other quarks in the nucleon. 
In inclusive DIS, the Operator Product Expansion 
applies and power corrections are determined by the matrix elements of
operators of higher twist (dimension-spin), which are related to 
multiparton configurations.

Such effects are inversely proportional to $Q^2$, and can
therefore be large at small $Q^2$. If they are not,
averages of the $F_2$ structure
function over a sufficient range in $x$ at moderate $Q^2$
are approximately the same as at high $Q^2$.
Notwithstanding, the dynamical origin of local duality, and thus the reason why
the higher-twist contributions, undoubtedly required to construct the coherent
nucleon resonances, tend to largely cancel on average, {\sl even} at momentum
transfers below 5 (GeV/c)$^2$, is still not understood \cite{dukerob,ji}.

\section{Local Duality at Low Momentum Transfer}

In Fig. 1 we show an overview of recent high-precision proton
resonance $F_2$ data at low $Q^2$ \cite{ioana2}. We also include data
from SLAC at $Q^2 <$ 0.3 (GeV/c)$^2$ \cite{stein}. For the former, the
systematic uncertainty is estimated to be 3.5\% \cite{ioana2}. For the
latter, due to uncertainties in absolute normalization and radiative
corrections, we estimate the systematic uncertainty to be better than 10\%.
The solid curves represent, for the different kinematics,
the single scaling curve defined by averaging {\sl all} nucleon resonance 
$F_2$ data, regardless of $Q^2, W^2$, as a function of $\xi$ \cite{ioana2}.
As one can see the individual spectra, at various $Q^2$, oscillate around this
single-curve parameterization. We emphasize that this is {\bf not} by construction,
as the parameterization, at any given value of $\xi$, is obtained from a
range of nucleon resonance data at variant values of $Q^2$ and $W^2$ (e.g.,
the second resonance bump could have always been below the scaling curve,
while the first above, etc.).
Our main observation is that 
apparently nature forces the oscillatory behavior of the various resonance
bumps around a scaling curve, which has a valence-like behavior.

This has been studied quantitatively in \cite{ioana2} where 
it was observed that the behavior of averaged nucleon resonance data at $\xi >$ 0.3,
corresponding to $Q^2 \ge$ 0.5 (GeV/c)$^2$ in the nucleon resonance region,
is indistinguishable from the $F_2$ DIS behavior, consistent with the findings of
Bloom and Gilman \cite{bg}.
The behavior of averaged nucleon resonance data for $\xi <$ 0.3, 
corresponding to $Q^2 <$ 0.5 (GeV/c)$^2$ in the nucleon
resonance region, mimics \cite{ioana1} $xF_3$ data obtained from averaging
neutrino and antineutrino DIS data \cite{conrad}.
The latter, to leading order in QCD, selects the difference of quark
and antiquark distribution functions, and is predominantly sensitive to
a valence quark only distribution.

Increasing from $Q^2 \approx$ 0.07 (GeV/c)$^2$ (Fig. 1a) to $Q^2 \approx$
3 (GeV/c)$^2$ (Figs. 1h,1i) the $F_2$ spectra change shape drastically.
The low $Q^2$ spectrum shows a predominant $N-\Delta$ transition
(we do not show the elastic peak, huge at this $Q^2$), and relatively
minor strength at larger energy transfers. This is not surprising, as at these
relatively small energy and four-momentum transfers one would expect to 
predominantly excite the valence quarks. At $Q^2$ = 3.0 (GeV/c)$^2$, on
the other hand, one sees that the prominent nucleon resonances are largely
reduced, and the inelastic background enhanced.
Furthermore, $F_2$ in the higher resonance regions is larger
than $F_2$ in the $N-\Delta$ transition region. Apparently, a swap of
strength has occurred between the various channels as a function of $Q^2$.

To further illustrate how the nucleon resonances seem to follow a valence-like
curve, we show in Fig. 2 the behavior of the $N-\Delta$
transition region
(here defined as 1.2 $< W^2 <$ 1.9 GeV$^2$) and the second resonance
region (defined as 1.9 $< W^2 <$ 2.5 GeV$^2$) for various $Q^2$ as a function
of $\xi$, in comparison with the global scaling curve defined in \cite{ioana1},
and used in Fig. 1.
As concluded in Refs. \cite{ioana1,ioana2}, it seems that the
nucleon resonances slide along one global scaling curve (note that the apparent
difference in scaling curve between Figs. 1 and 2 reflects only the conversion
from $W^2$ to $\xi$, for fixed $Q^2$).
One can see that, if nature forces the oscillatory behavior around a
global scaling curve even at low $Q^2$, the resonance excitation strengths will
necessarily grow in the region below $\xi \approx$ 0.25 where
the maximum of the global scaling curve occurs, and subsequently decrease once
the maximum of the global scaling
curve has been crossed.
Compare, for instance,
the behavior of the $N-\Delta$ transition region with that
of the larger-mass resonance regions: at $Q^2$ = 0.45 (GeV/c)$^2$
(solid circles in Fig. 2 (top), and Fig. 1c) the $N-\Delta$ transition region
strength is large being at about the maximum of the scaling curve.
Its strength, as for all $Q^2 <$ 0.45 (GeV/c)$^2$, is also larger than the
higher-mass resonance regions which lie at lower $\xi$ for
the same $Q^2$. On the other hand, for $Q^2 \approx$ 3.0 (GeV/c)$^2$
(open circles in Fig. 2(top), and Fig. 1h) the $N-\Delta$ transition region
strength is small because it is positioned at large $\xi$, and smaller than the
higher-mass resonance regions that lie at lower values of $\xi$, but have
crossed the maximum of the scaling curve.
The smooth curve, to which the nucleon resonance regions tend,
determines the momentum transfer dependence of the various
nucleon resonance regions, forcing the nucleon elastic and transition form
factors to scale like $Q^{-4}$ \cite{stoler} at relatively small $Q^2$,
resembling the $Q^{-4}$ scaling as predicted by QCD counting rules.

We now ask the question: Can the observed behavior of the averaged nucleon
resonance spectra be explained within pQCD?

In Fig. 3, we show a compilation of the world's electron-proton scattering
data for the $F_2$ structure function at low $Q^2$. The deep inelastic
($W^2 >$ 4 GeV$^2$) data originate from
SLAC \cite{slac}, CERN (NMC) \cite{nmc97}, FNAL (E665) \cite{e665}, and DESY
(H1,ZEUS) \cite{h196,zeus95,zeus96,zeus97,zeus98,zeus00}. As before we include data
in the proton resonance region from SLAC \cite{stein} and JLab \cite{ioana1}.
The solid curves indicate the next-to-leading order parameterizations
of Gl\"uck, Reya, and Vogt (GRV) \cite{grv95,grv98}, which use input parton
distributions starting from very low $Q^2$ values.
In the third panel down of Fig. 3, the GRV calculation is
the GRV input distribution at $Q^2$ = 0.4 (GeV/c)$^2$ \cite{grv98},
without any evolution, {\sl neglecting sea and gluon contributions}.
As one can see, the proton resonance region data for $Q^2$ = 3.1 (GeV/c)$^2$
and $Q^2$ = 0.9 (GeV/c)$^2$ smoothly join the deep inelastic data, and agree
well with the GRV next-to-leading order calculations, exhibiting the
local duality witnessed by Bloom and Gilman \cite{bg}.

Turning our attention to the bottom two panels of Fig. 3, the low $Q^2$
$F_2$ data, we are only left with the recent DESY data \cite{zeus97,zeus98},
some sparse FNAL data \cite{e665}, and the nucleon resonance data
\cite{stein,ioana1}. The DESY data exhibit the well-known collapse of
the proton structure functions at (very) small $x$. One interpretation of
this effect is that, at these small $Q^2$, one sees  a smooth transition from DIS
to the real photon point at $Q^2$ = 0 \cite{zeus98,zeus00,donnach}.
Gauge invariance requires that, for consistency near $Q^2$ = 0, the structure
function $F_2$ for inelastic channels
must vanish like $Q^2 \sigma(\gamma p)/(4\pi^2 \alpha_{em})$ \cite{donnach}.

We emphasize here the difference in reaching the low $Q^2$ region for
the various values of $x$. In Fig. 4 we distinguish among three different
limits (indicated by the three arrows in the figure). 
For the DESY experiments, low $Q^2$ is established
at small $x$ by having a large ($\approx$ constant)
amount of energy transfer $\nu$. For this reason,
it is expected that this region exhibits similar characteristics as the
parton model.
For the JLab/SLAC experiments at $x \approx$ 0.1, one reaches
low $Q^2$ at relatively small energy transfers.
If the limit is taken at {\em fixed} $x$, $x \approx$ 0.1, 
the resonance data do not
exhibit such a drastic collapse with $Q^2$ as observed at HERA, 
and they stay fairly constant. In fact,
they still seem to oscillate around one $Q^2$-independent global curve \cite{ioana1},
informing us that the $Q^2$ dependence of the larger $x$ nucleon resonance data
is rather shallow. It is this behavior that makes the scaling curve of
Figs. [1,2], defined by the world's nucleon resonance $F_2$ data,
look ``valence-like" \cite{ioana1}.
If, instead, one takes the limit at a fixed $W^2$, by restricting
oneself always to the nucleon resonance region ($M^2 < W^2 \le$ 4 GeV$^2$),
then one can see that the nucleon
resonances ``slide" down the $x$ scale to lower $x$ for lower $Q^2$
({\it e.g.} see the constant $W$ = 2 GeV arrow in Fig. 4), where their strength
dies out as a function of $Q^2$.

\section{Moments of $F_2^p$}

We construct the experimental moments of the structure function, $F_2$,
for the $Q^2$ range up to 10 (GeV/c)$^2$.
The Cornwall-Norton moments are defined as
\begin{equation}
M_n(Q^2) = {\int_0^{x_{thr}}} dx x^{n-2} F_2(x,Q^2),
\label{CN}
\end{equation}
and the Nachtmann moments as
\begin{eqnarray}
M_n(Q^2) = {{\int_0^{x_{thr}}} dx {{\xi^{n+1}} \over {x^3}}}
\nonumber \\
& & \hspace{-25mm}
\bigg[{{3+3(n+1)r+n(n+2)r^2} \over {(n+2)(n+3)}}\bigg]
\nu W_2(x,Q^2).
\end{eqnarray}
Here, $r$ = (1 + 4$M^2$$x^2$/$Q^2$)$^{1/2}$,
and $x_{thr}$ is Bjorken $x$ for pion threshold. We add to these integrals
the elastic contribution, at $x$ = 1, where
\begin{equation}
{\nu W_2(x,Q^2) = \delta(1-x) 
{{\bigg(G_E^2(Q^2) + {{Q^2} \over {4M^2}} G_M^2(Q^2)\bigg)}
\over {\bigg(1 + {{Q^2}\over{4M^2}}\bigg)}}.}
\end{equation}
$G_E$ ($G_M$) is the proton electric (magnetic) form factor.
For the proton form factors, we use
a fit to the world's data by Bosted \cite{bosted}.

To obtain the inelastic contributions, we integrate data like those shown in
Figs. 1 and 3.
Apart from the data shown in Fig. 1, we have added $Q^2$ points where
additional data were available. For $Q^2 <$ 0.6 (GeV/c)$^2$, we have
constrained our search to elastic and nucleon resonance data.
For 0.6 $< Q^2 <$ 4 (GeV/c)$^2$, we have used nucleon resonance
data in combination with deep inelastic data, whereas for $Q^2 >$ 4 (GeV/c)$^2$
we have constructed the moments utilizing both deep inelastic and
nucleon resonance models, similar as in Ref. \cite{twistpaper}.
For the smallest values of $Q^2$ ($<$ 0.6 (GeV/c)$^2$),
we assume a constant value of $F_2$ below $x$ for $W^2$ = 4.0 GeV$^2$,
as only sparse data exists.
As one can see from Fig. 3, this may not be a bad
approximation for $Q^2 <$ 0.6 (GeV/c)$^2$, especially since the nucleon
resonance region data extend down to $x \le$ 0.1, and the integration area
below $x$ = 0.1 is expected to be small only. To judge the uncertainty in
this procedure, we have also integrated the $Q^2 \approx$ 0.2 (GeV/c)$^2$
data starting at $W^2$ = 9.0 GeV$^2$ (rather than $W^2$ = 4.0 GeV$^2$).
This changes the second moment by less than 3\%.
Lastly, in some cases, we used a model to construct data at fixed $Q^2$, rather
than allowing for the small range of $Q^2$ in the data.
This effect on the second moments was found to be small, $<$ 3\%, and far
smaller for the higher moments.
Thus, we believe the total uncertainty in the moments we calculate to be
less than 5\%.
We show the values for the second, fourth, sixth, and eigth
Cornwall-Norton (top) and Nachtmann (bottom) moments of the proton,
extracted from the world's data, including
deep inelastic, nucleon resonance, and elastic data, as described above,
in Fig. 5. Similarly, Tables I and II list the numerical values of the
moments, with the elastic contribution to each separately given.

As expected, the elastic contribution dominates the moments
at the lowest $Q^2$. Note that the Cornwall-Norton
moments will become unity, i.e. the proton charge squared, at $Q^2$ = 0, 
whereas the Nachtmann moments will vanish
at $Q^2$ = 0, as can readily be seen from Eqn. 3. 
This can be attributed 
to the fact that, with respect to Bjorken $x$, the Nachtmann scaling variable
$\xi$ correctly takes into account the finite proton mass scale \cite{nacht},
but does not account for any other significant mass scale
(like the quark masses).
As the interpretation of the Cornwall-Norton moments in the $Q^2 <$ 1
(GeV/c)$^2$ region seems more intuitive,
and we are interested here in the low-$Q^2$ behavior of the moments,
we will concentrate on these moments in the remainder of the discussions
in this work.
To emphasize that there is indeed not much difference between the
Cornwall-Norton and Nachtmann moments if one concentrates on the
low-$Q^2$ region where the elastic contribution turns dominant, Fig. 6
graphically displays the relative contribution
of the elastic channel to the total moment for both Cornwall-Norton (top)
and Nachtmann (bottom) moments, for $n$ = 2 (solid circles), $n$ = 4 (squares),
$n$ = 6 (triangles) and $n$ = 8 (stars), from Tables I and II.
Nonetheless, as the benefit of the Nachtmann moments is to push an analysis in
terms of an Operator-Product Expansion to lower values of $Q^2$, taking
correctly target-mass effects into account, we show everywhere similar figures,
for comparison, for the Nachtmann moments. Note that one can argue that
the relative contribution of the elastic grows slower for
$Q^2 \rightarrow$ 0 if one uses the Nachtmann moments.

First, let us revisit the large $Q^2$, $Q^2 > 10$ (GeV/c)$^2$, behavior of the 
moments. At asymptotically large $Q^2$, one can write 
the non-singlet moments $M_n^{NS}$($Q^2$) at Leading Order (LO)
\footnote
{NLO corrections are important in general 
and will be discussed in detail in \protect{\cite{liuti}}. 
Their discussion is, however, not essential for discussing
the points raised in the present paper.}
in the perturbative expansion, as 
\cite{bur,dukerob}  
\begin{equation}
M_n^{NS}(Q^2) = A_n (ln(Q^2/\Lambda^2))^{-1/d_n^{NS}},
\end{equation}
where $\Lambda$ is the QCD scale parameter and
\begin{equation}
d_n^{NS} = \gamma_{\circ,n}^{NS}/2\beta_\circ,
\end{equation}
where $\beta_\circ$ = 11 - (2/3)$N_f$, with $N_f$ the number of
flavors, and ${\gamma_{\circ,n}}^{NS}$ the leading-order non-singlet
anomalous dimensions numerically specified in \cite{bur}.
To circumvent the requirement of non-singlet moments we highlight
in Fig. 7 the $n$ = 4 moment:  
although we use the same $F_2$ structure function data for
all moments, the weighting with $x^{n-2}$ in these moments will emphasize the
large-$x$ region, at higher $n$, and thus approximate a non-singlet moment.
On the top of Fig. 7 we show
the Cornwall-Norton moment, on the bottom the Nachtmann moment (each to
the power $-1/d_n^{NS}$). We show the data in a log-log plot,
in order to encompass also the low-$Q^2$ region.
The moments are shown both with (stars) and without
(open circles) the elastic contribution included.
 
The dashed curves exhibit a fit to the data,
from \cite{twistpaper},
limited to $Q^2 >$ 20 (GeV/c)$^2$ to minimize the effect of higher order
corrections,
in the form $[M_n^{NS}]^{(-1/d_n^{NS})} \approx P_1 ln(Q^2/\Lambda^2)$. 
In \cite{twistpaper}, this fit gives
$P_1$ = 27.46 (27.05) $\pm$ 0.25 (0.24) and $\Lambda$ = 250 MeV,
for the Cornwall-Norton (Nachtmann) moment, rendering the
expected logarithmic scaling behavior in QCD at asymptotic $Q^2$.
The dot-dashed (dotted) curves exhibit a similar fit in the form
$(P_1 + P_2/Q^2 + P_3/Q^4)^{-1/d_n} ln(Q^2/\Lambda^2)$ 
down to $Q^2$ = 2.0 (GeV/c)$^2$ from the same Reference \cite{twistpaper},
thus taking into account power corrections of order 1/$Q^2$ and 1/$Q^4$
(1/$Q^2$ only).
Numerical values for the $1/Q^2$ and $1/Q^4$
coefficients are
$P_2$ = 0.33 $\pm$ 0.04 (0.33 $\pm$ 0.04) and $P_3$ = 4.69 $\pm$ 0.19
(1.61 $\pm$ 0.15) for the Cornwall-Norton (Nachtmann) moment (see also
the caption of Fig. 7).
One can easily verify from Fig. 7 that the magnitude
of the $P_3$ coefficient is in this case dominated by the inclusion of
the elastic contribution. Similarly, the Nachtmann ($n = 4$) moment analysis
gives a drastically different value for $P_3$ from the Cornwall-Norton ($n=4$)
moment analysis mainly due to the different contribution from the elastic.

In order to illustrate the relative  
strength of the elastic contribution compared to other $W^2$ regions, we show
in Fig. 8 the second (Fig. 8a), fourth (8b), sixth (8c) and eigth (8d)
Cornwall-Norton moments for $Q^2 <$ 5 (GeV/c)$^2$,
separated in the elastic contribution (squares, due to
our choice of vertical scale sometimes only visible at the higher $Q^2$), the
contribution of the $N-\Delta$ transition region (triangles,
1.2 $< W^2 <$ 1.9 GeV$^2$), of the second resonance region
(open circles, 1.9 $< W^2 <$ 2.5 GeV$^2$) and of the
``deep inelastic" region (stars, $W^2 >$ 4 GeV$^2$). The total moment is
given by the solid circles, and the curves connect the various data to guide
the eye. The 
chosen finite $W^2$ regions will start contributing to the moments at low $Q^2$,
recovering part of the loss of strength due to the fall-off of the elastic
contribution, and then also die off, as the resonances move to the larger
$\xi$ side of the scaling curve. The contribution of the $W^2 >$ 4 GeV$^2$
region does not die off, as this is not a finite $W^2$ region, so higher-$W^2$
resonances and/or higher-$W^2$ inelastic background start becoming important
with increasing $Q^2$, eventually yielding the logarithmic behavior of the
moments prescribed by QCD. As evidenced by the difference between the
$W^2 >$ 4 GeV$^2$ contribution and the total moment,
the contribution of the
region of $W^2 <$ 4 GeV$^2$ is non-negligible up to $Q^2 \approx$
5 (GeV/c)$^2$, even for the second moment.
Similar remarks hold for the various Nachtmann moments,
shown in Fig. 9.  

We note here also that the behavior of the second Cornwall-Norton $F_2$ moment
we extract is very similar to the behavior in the second moment of
the spin-dependent $g_1$ structure function \cite{jimel}. Presently we only
have sparse $g_1$ data in the nucleon resonance region, for
0.1 $< Q^2 <$ 5 (GeV/c)$^2$, such that we can not verify precisely whether
the spin-dependent nucleon resonance data tend to oscillate around a
similar smooth curve. However, the limited data are not inconsistent with
such a behavior \cite{e143}. Also, we presently do not have enough data
for the longitudinal structure function $F_L$ to verify a similar onset of
duality \cite{carmuk}, although sparse hints do exist in the present world's
data \cite{ent}.

\section{Discussion of Results}

Our findings, that the moments of $F_2$ show a smooth transition from
DIS down to $Q^2 \approx$ 0 (GeV/c)$^2$, and that the
nucleon resonances tend to oscillate around one smooth curve, support the
findings of Ref. \cite{bodek,kataev,liuti} that higher-twist effects are small
if one looks at the low-$Q^2$ behavior of $F_2$ for $Q^2 \simeq$ 1 (GeV/c)$^2$.
The dynamical process of local duality dictates minimal
$Q^2$ dependence of $F_2$ at small $Q^2$; in terms of pQCD 
this can be explained if the higher-twist effects
are reduced {\sl on average} in the nucleon resonance region \cite{gp}.
Nonetheless, higher-twist effects must be responsible for the nucleon
resonances themselves. The results for the lower moments of $F_2$,
presented here, show a forced transition from the elastic point to the
large $Q^2$ limit, supported by the oscillations of the nucleon
resonance region around one smooth curve at low $Q^2$. This smooth curve
resembles the deep inelastic data at $Q^2 \simeq$ 1 (GeV/c)$^2$, and
higher-twist effects continue to be small from there on.

The extension of a pQCD analysis to very low values of $Q^2$,
{\it i.e.} $Q^2 \rightarrow \Lambda^2$, is of doubtful predictivity. 
Here even the low $n$ moments are mainly sensitive to the inclusion of
the elastic contribution (as this contributes already close to 10\% 
to the $n=2$ moment at $Q^2$ = 2 (GeV/c)$^2$):
a picture in terms of non-interacting quark and gluon degrees
of freedom is clearly no longer tenable.
Similarly the most plausible interpretation of the behavior observed for the
dynamical power corrections,  which should represent the initial signature
of NP effects overshadowing the perturbative expansion, would in fact be
that their contribution in the 1/$Q^2$ expansion is enhanced at decreasing
values of $Q^2$. However, this contribution can be damped on average
if coefficients of different powers are large but with opposite signs,
thus causing the cancellation required to make the moment
a slowly varying function of $Q^2$ only. These cancellations are evidently
of NP origin and are at low $Q^2$ not easily associated to partonic degrees of 
freedom in the multiparton correlations.

One has to be careful making further concrete statements. E.g., it is
often remarked that duality obviously does not work universally, as the
$n$ = 2 Cornwall-Norton moment at $Q^2 \rightarrow 0$ equals the coherent sum
of the quark charges squared, which can not equal the parton model expectation
in the Bjorken limit. First of all, $F_2$ is not purely transverse, which casts
doubt on drawing too definite a conclusion at low values of $Q^2$.
Results of an experiment to measure $R = \sigma_L/\sigma_T$ in the nucleon
resonance region are needed and forthcoming \cite{Rmeas}.
More importantly, the moment analysis will at lower and lower
$Q^2$ be sensitive to a smaller and smaller region in $\xi$, as the 
inelastic region and nucleon
resonances move to lower $\xi$ and their absolute contribution to the moments
diminishes. Even in a world without elastic electron-proton scattering,
one would, at the lowest $Q^2$, be mainly sensitive to a single resonance
transition region, in this case the $N-\Delta$ transition region.
This is illustrated in Fig. 10, where we repeat Fig. 5 for a world without
elastic electron-proton scattering. It can easily be
seen (Figs. 1, 3, 8, and 9) that the $N-\Delta$ transition contribution,
defined by the region 1.2 $< W^2 < 1.9$ GeV$^2$, becomes more and more
dominant at low $Q^2$. Thus, in this world a moment analysis at low $Q^2$
would become dominated by the coherent $N-\Delta$ process. Still, {\sl
on average} the $N-\Delta$ transition region seems to obey duality, in that
it oscillates around a similar smooth scaling curve. This is not inconsistent
with the findings of \cite{dgp}, the discrete resonance transitions are
reminiscent of large higher-twist effects, but on average still seem to
cancel to large extent (although quantitatively perhaps not as well as at
higher $Q^2$). The above example is not dissimilar to stating that duality
will not work on top of any given resonance peak. A region over which to
average is always required.

Furthermore, if one neglects the
elastic channel, one will at low $Q^2$ be mainly sensitive to
the imposed constraint by gauge invariance that the structure
function $F_2$ must behave like $Q^2 \sigma(\gamma p)/(4\pi^2 \alpha_{em})$
\cite{donnach}.
At the values of $x$ where the nucleon resonances are visible
at low $Q^2$ in Fig. 3 (e.g. $x \approx$ 0.1), the $F_2$ structure function
does {\bf not} linearly vanish with $Q^2$ yet, as shown in \cite{ioana1}.
Thus, although the $F_2$ strength in the nucleon resonance region has to
disappear linearly with $Q^2$ below some $Q_0^2$, one can argue that
the behavior of the data is not reflecting this $Q^2 < Q_0^2$ expectation yet.
This indicates that the oscillations the nucleon resonances exhibit
around a smooth curve, even down to $Q^2 \approx$ 0.1 (GeV/c)$^2$, is
non-trivial.
As the low-$Q^2$ $F_2$ data below $W^2$ = 4 GeV$^2$
predominantly consists of excited nucleon resonances and hardly contributions
from inelastic non-resonant processes, one can argue that such a smooth
curve must be close to a curve
consisting of valence strength only. In fact, the $Q^2$ dependence of the
integrated valence quark strength in the GRV model \cite{grv95,grv98} is close
to the $Q^2$ dependence of the second Cornwall-Norton moment of $F_2$.
However, this $Q^2$ dependence is predominantly due to the inclusion of the
elastic channel.
Thus, for a picture such as the GRV model to be valid,
there must be a separate $Q^2$ dependence for
the vanishing of the large-$x$ strength at small $Q^2$ (governed by the
nucleon resonances) and the growth of the small-$x$ sea.

Such ideas are very similar as to what has been observed in hadron-hadron
scattering \cite{harari}. Here, a generalization of the duality picture was
introduced, in which resonances were dual to non-diffractive Regge pole
exchanges, with the non-resonant contributions dual to Pomeron exchange.
Within QCD, this corresponds to a picture where resonances are dual to
valence quarks, while the non-resonant background is dual to sea quarks.
This supports the importance of additional detailed studies of electron
scattering in the nucleon resonance region, in a wide range of $x$ and $Q^2$.

\section{Conclusions}

We show that the world's data on $F_2$, down to $Q^2 \simeq$
1 (GeV/c)$^2$, are reasonably well described within pQCD. This includes
the nucleon resonance data, which average to an approximate scaling curve,
due to local duality. 
Down to $Q^2 \simeq$ 0.1 (GeV/c)$^2$, the nucleon
resonance data still tend to average to such a curve.
The contribution of the nucleon resonances to the lower moments of $F_2$ 
dies out at very small $Q^2$ as they have moved to smaller Bjorken $x$.
Instead, the moments below $Q^2 \approx$ 1 (GeV/c)$^2$ are
dominated by the elastic contribution at large $x$.
Therefore, a pQCD-based analysis of the low-$Q^2$ moments of $F_2$ will
render coefficients of the higher-twist terms
which are predominantly due to the elastic contribution.
Local duality seems to hold down to at least $Q^2 \approx 0.5$ (GeV/c)$^2$ 
and to prescribe the transition from the kinematic region
dominated by the elastic contribution
to the region dominated by deep inelastic scattering.

\medskip
This work was supported in part by the U.S. Department
of Energy under Grant No. DE-FG02-95ER40901, and
the National Science Foundation under Grant No. HRD-9633750.
The authors express gratitude to the
Jefferson Lab Theory Group for many useful discussions.
CEK acknowledges the support of an NSF Early Faculty Career
Development Grant.

\begin{table}
\caption{Cornwall-Norton Moments for $n$ = 2, 4, 6, and 8
at 0.15 $\le Q^2 \le$ 4.3 (GeV/c)$^2$, as extracted from the data (see text).
The elastic contribution is given as a separate column.
The uncertainties of the total moments are smaller than 5\%.}
\begin{tabular}{|c|c|cccc|}
$Q^2$ (GeV/c)$^2$ & elastic & $n=2$ & $n=4$ & $n=6$ & $n=8$ \\
\hline
\hline
0.15  & 0.592 & 0.652 & 0.594 & 0.592 & 0.592 \\
0.20  & 0.504 & 0.584 & 0.508 & 0.505 & 0.504 \\
0.45  & 0.249 & 0.379 & 0.261 & 0.251 & 0.250 \\
0.55  & 0.195 & 0.341 & 0.210 & 0.198 & 0.196 \\
0.85  & 0.103 & 0.278 & 0.122 & 0.107 & 0.104 \\
0.94  & 0.087 & 0.264 & 0.107 & 0.092 & 0.088 \\
1.40  & 0.040 & 0.231 & 0.064 & 0.047 & 0.043 \\
1.70  & 0.026 & 0.219 & 0.051 & 0.034 & 0.029 \\
2.40  & 0.011 & 0.203 & 0.036 & 0.019 & 0.014 \\
3.00  & 0.006 & 0.196 & 0.030 & 0.013 & 0.009 \\
3.30  & 0.005 & 0.192 & 0.028 & 0.012 & 0.008 \\
4.30  & 0.002 & 0.184 & 0.023 & 0.008 & 0.005 \\
\end{tabular}
\label{table:CNmoments}
\end{table}

\begin{table}
\caption{Nachtmann Moments for $n$ = 2, 4, 6, and 8
at 0.15 $\le Q^2 \le$ 4.3 (GeV/c)$^2$, as extracted from the data (see text).
The elastic contributions, different for each $n$, are given as a separate
entity in the columns.
The uncertainties of the total moments are smaller than 5\%, and the numbers
quoted can be used to this precision.}
\begin{tabular}{|c|cc|cc|cc|cc|}
$Q^2$ & $n=2$ & & $n=4$ & & $n=6$ & & $n=8$ & \\
      & elas. & total & elas. & total & elas. & total & elas. & total \\ 
\hline
\hline
0.15 & 0.274 & 0.333 & 0.040 & 0.041 & 0.0051 & 0.0052 & 0.0006 & 0.0006 \\
0.20 & 0.256 & 0.322 & 0.047 & 0.049 & 0.0074 & 0.0075 & 0.0011 & 0.0011 \\
0.45 & 0.160 & 0.281 & 0.050 & 0.057 & 0.0139 & 0.0146 & 0.0037 & 0.0038 \\
0.55 & 0.131 & 0.268 & 0.046 & 0.056 & 0.0146 & 0.0157 & 0.0045 & 0.0046 \\
0.85 & 0.076 & 0.243 & 0.034 & 0.047 & 0.0137 & 0.0157 & 0.0054 & 0.0057 \\
0.94 & 0.065 & 0.235 & 0.030 & 0.045 & 0.0130 & 0.0154 & 0.0054 & 0.0058 \\
1.40 & 0.032 & 0.217 & 0.018 & 0.036 & 0.0093 & 0.0128 & 0.0047 & 0.0055 \\
1.70 & 0.022 & 0.209 & 0.013 & 0.033 & 0.0073 & 0.0114 & 0.0040 & 0.0051 \\
2.40 & 0.010 & 0.197 & 0.007 & 0.027 & 0.0041 & 0.0089 & 0.0026 & 0.0040 \\
3.00 & 0.005 & 0.192 & 0.004 & 0.024 & 0.0026 & 0.0075 & 0.0018 & 0.0033 \\
3.30 & 0.004 & 0.189 & 0.003 & 0.023 & 0.0021 & 0.0071 & 0.0015 & 0.0031 \\
4.30 & 0.002 & 0.181 & 0.001 & 0.020 & 0.0011 & 0.0058 & 0.0008 & 0.0024 \\
\end{tabular}
\label{table:Nachtmannmoments}
\end{table}

\begin{figure}
\begin{center}
\epsfxsize=2.75in
\epsfysize=2.75in
\epsffile{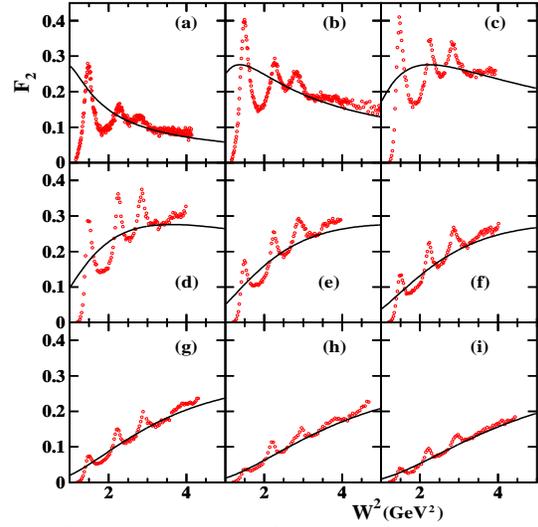}
\caption{$F_2$ Spectrum in the nucleon resonance region
as a function of $W^2$, for values of
$Q^2$ = 0.07 (a), 0.20 (b), 0.45 (c), 0.85 (d), 1.40 (e), 1.70 (f),
2.40 (g), 3.00 (h) and 3.30 (i) (GeV/c)$^2$.
We have superimposed the results from the scaling curve from Ref.
\protect\cite{ioana2}, to illustrate the behavior of the nucleon
resonance region with increasing $Q^2$.}
\end{center}
\end{figure}

\begin{figure}
\begin{center}
\epsfxsize=2.75in
\epsfysize=2.75in
\epsffile{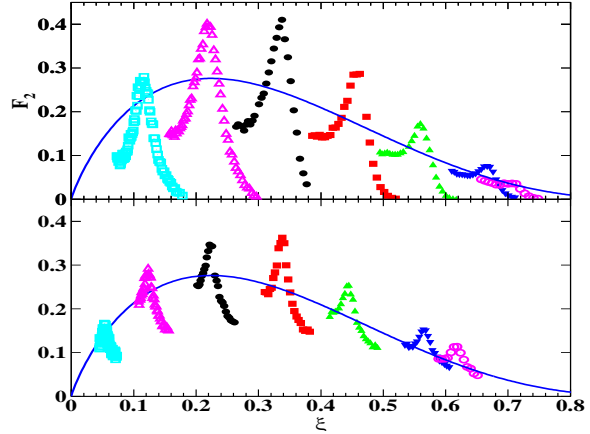}
\caption{$F_2$ data for the regions 1.2 $< W^2 <$ 1.9 (top) and
1.9 $< W^2 <$ 2.5 (bottom) GeV$^2$, as a function of Nachtmann $\xi$.
Data are shown for $Q^2$ = 0.07, 0.20, 0.45, 0.85, 1.4, 2.4, and 3.0
(GeV/c)$^2$ (left to right), respectively.
The solid curve represents the scaling curve, determined by averaging
{\sl all} nucleon resonance data \protect\cite{ioana2}.}
\end{center}
\end{figure}

\begin{figure}
\begin{center}
\epsfxsize=3.25in
\epsfysize=5.25in
\epsffile{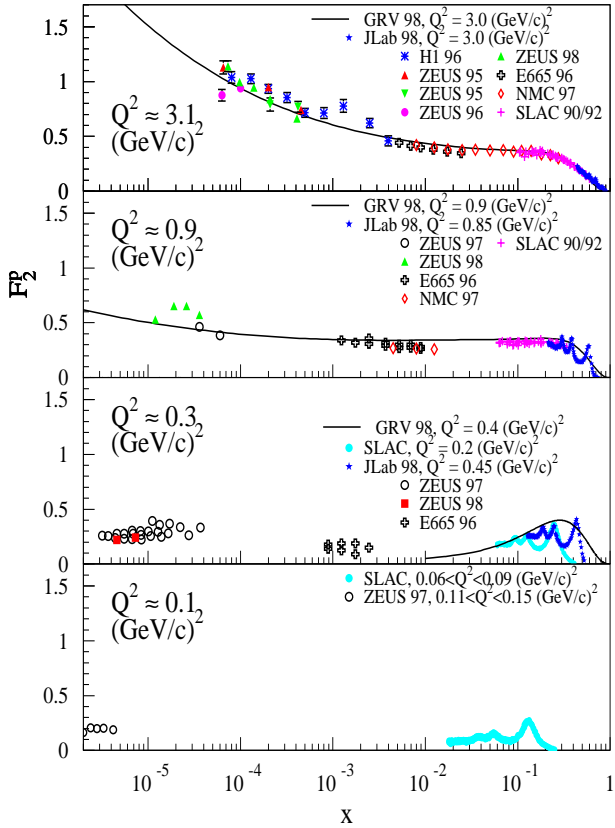}
\caption{$F_2$ as a function of $x$ for four values of $Q^2$,
with a logarithmic $x$ scale. The symbols indicate various experiments,
as cited in the text. The solid curves in the top two panels represent the
calculated distributions from the GRV collaboration \protect\cite{grv95,grv98},
evolved from $Q^2$ = 0.4 (GeV/c)$^2$. The solid curve in the third panel
represents the input distribution itself.}
\end{center}
\end{figure}

\begin{figure}
\begin{center}
\epsfxsize=3.25in
\epsfysize=3.25in
\epsffile{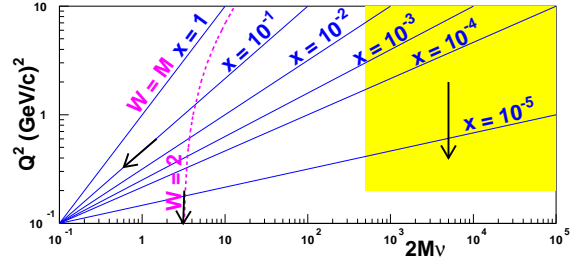}
\caption{Kinematics region of the world's $F_2$ structure function data,
for $Q^2 <$ 10 (GeV/c)$^2$. Thin lines are for fixed Bjorken $x$. The dashed
line is for fixed $W$ = 2 GeV, indicating the border of the region typically
associated with the nucleon resonances (the other border being fixed $W$ = $M$,
or $x$ = 1).
The shaded area exhibits the kinematics region of the recent DESY measurements
\protect\cite{h196,zeus95,zeus96,zeus97,zeus98}.
The thick arrows indicate various manners in which one can reach the
limit $Q^2 \rightarrow$ 0, at fixed $x$, fixed $W$, or fixed $\nu$.}
\end{center}
\end{figure}

\begin{figure}
\begin{center}
\epsfxsize=3.25in
\epsfysize=2.75in
\epsffile{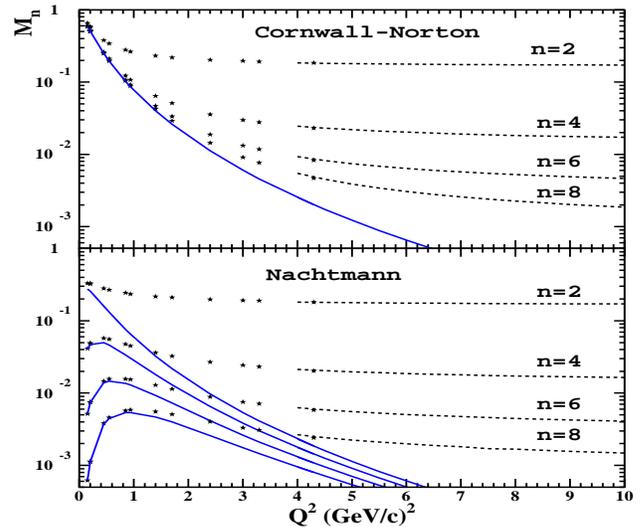}
\caption{Cornwall-Norton moments (top) and Nachtmann moments (bottom)
extracted from the world's electron-proton scattering data,
for $n$ = 2, 4, 6, and 8.
The solid lines indicate the elastic contribution.
At low $Q^2$ ($\le$ 4.3 (GeV/c)$^2$) the moments (stars) are directly
constructed from the world's electron-proton $F_2$ database (see text).
At larger $Q^2$, the moments have been extracted from appropriate fits to the
world's data on inclusive scattering to both the nucleon resonance and deep
inelastic regions (dashed lines) \protect\cite{twistpaper}.}
\end{center}
\end{figure}

\begin{figure}
\begin{center}
\epsfxsize=3.25in
\epsfysize=2.75in
\epsffile{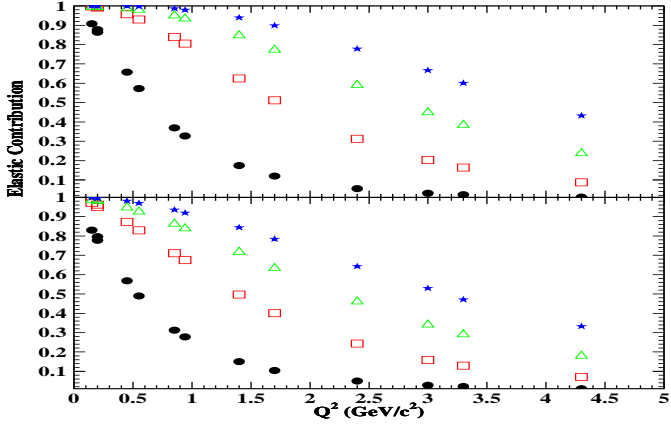}
\caption{Contribution of the elastic channel to the
Cornwall-Norton moments (top) and Nachtmann moments (bottom)
extracted from the world's electron-proton scattering data,
for $n$ = 2 (solid circles), 4 (squares), 6 (triangles), and 8 (stars),
up to $Q^2$ = 5 (GeV/c)$^2$. The data are from Tables I and II.}
\end{center}
\end{figure}

\begin{figure}
\begin{center}
\epsfxsize=3.25in
\epsfysize=3.75in
\epsffile{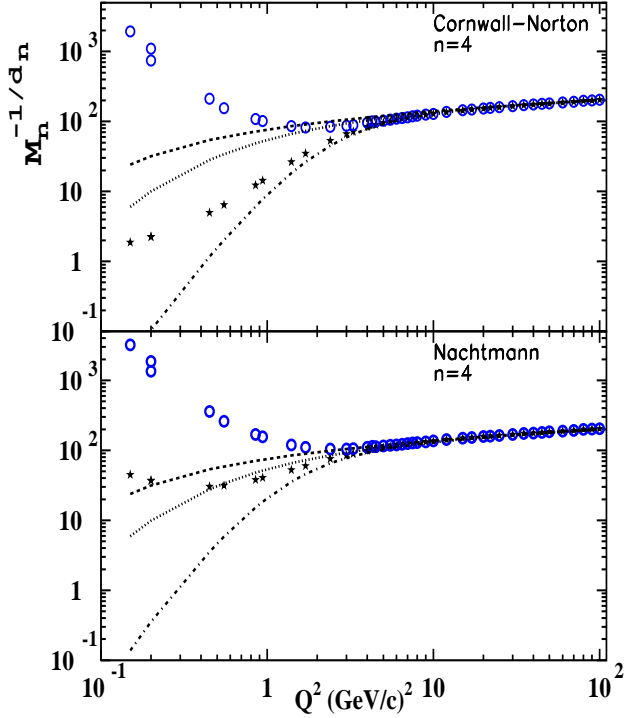}
\caption{Log-log plot of the $n$ = 4 Cornwall-Norton (top) and Nachtmann
(bottom) $M_n$($Q^2$)$^{-1/d_n}$ moment as a function of $Q^2$, where
$d_n$ = ${\gamma_\circ}^n/2\beta_\circ$ \protect\cite{roberts}.
Stars (open circles) do (do not) include the elastic contribution.
In the top plot, the dashed curve is a fit to the $n$ = 4 moment in the form
$P_1 ln(Q^2/\Lambda^2)$, from \protect\cite{twistpaper},
with $P_1$ = 27.46 ($\pm$ 0.25), and $\Lambda$ = 250 MeV.
Similarly, the dotted curve uses a form $(P_1 + P_2/Q^2) ln(Q^2/\Lambda^2)$,
with $P_2$ = 0.33 ($\pm$ 0.04), and the dot-dashed curve uses a form
$(P_1 + P_2/Q^2 + P_3/Q^4) ln(Q^2/\Lambda^2)$, with $P_3$ = 4.69 ($\pm$ 0.19).
In the bottom plot, the fit parameters are $P_1$ = 27.05 ($\pm$ 0.24),
$P_2$ = 0.33 ($\pm$ 0.04), and $P_3$ = 1.61 ($\pm$ 0.15), respectively
\protect\cite{twistpaper}.}
\end{center}
\end{figure}

\begin{figure}
\begin{center}
\epsfxsize=3.25in
\epsfysize=3.75in
\epsffile{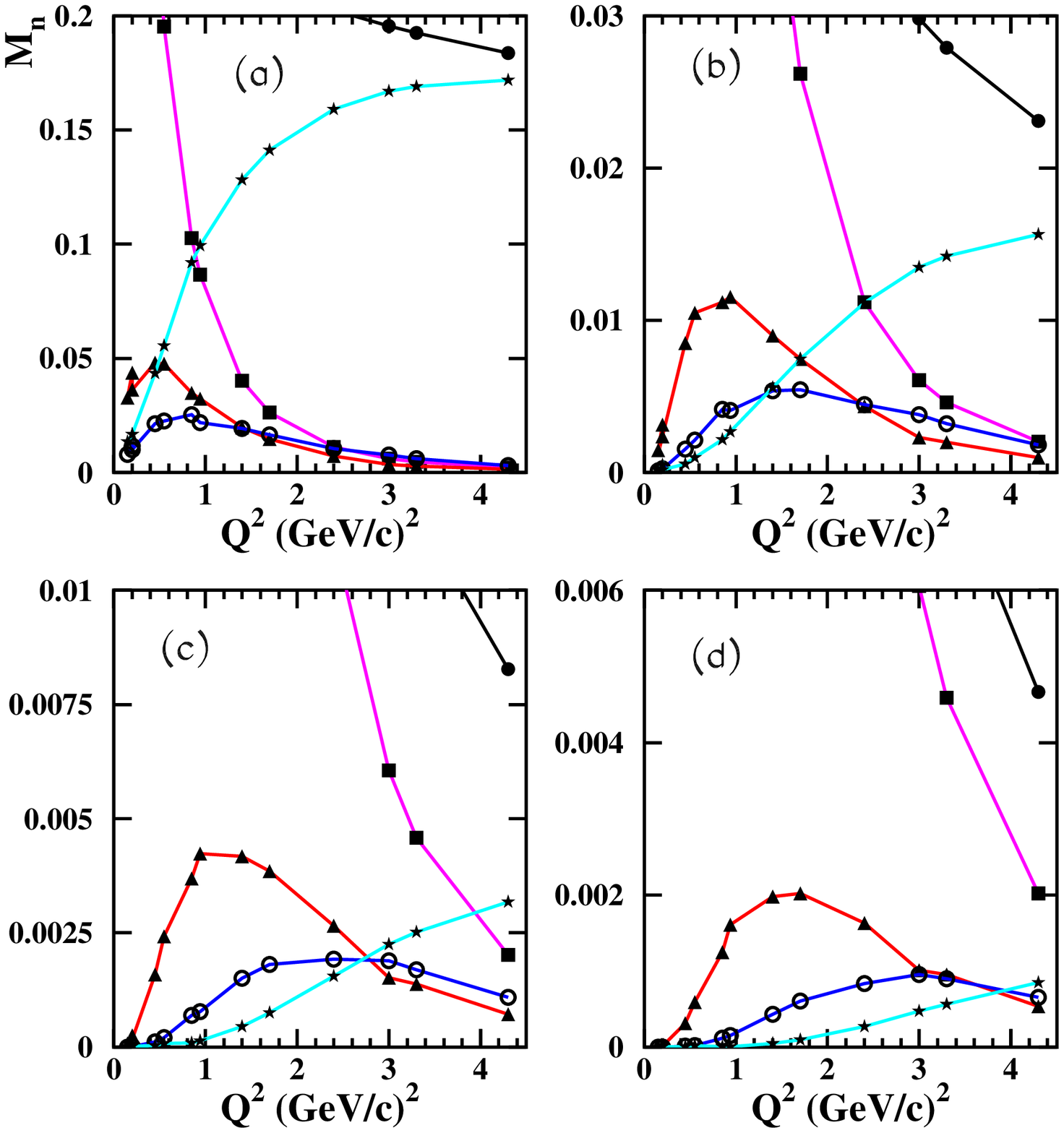}
\caption{Second (a), fourth (b), sixth (c) and eigth (d) Cornwall-Norton
moments. Contributions due to the elastic peak (squares), the regions
1.2 $< W^2 <$ 1.9 GeV$^2$ (triangles),
1.9 $< W^2 <$ 2.5 GeV$^2$ (open circles), and
$W^2 >$ 4 GeV$^2$ (stars) are separately shown, in combination with
the total moment (solid circles), as a function of the momentum transfer.
Curves connect the various data, and are to guide the eye only.}
\end{center}
\end{figure}

\begin{figure}
\begin{center}
\epsfxsize=3.25in
\epsfysize=3.75in
\epsffile{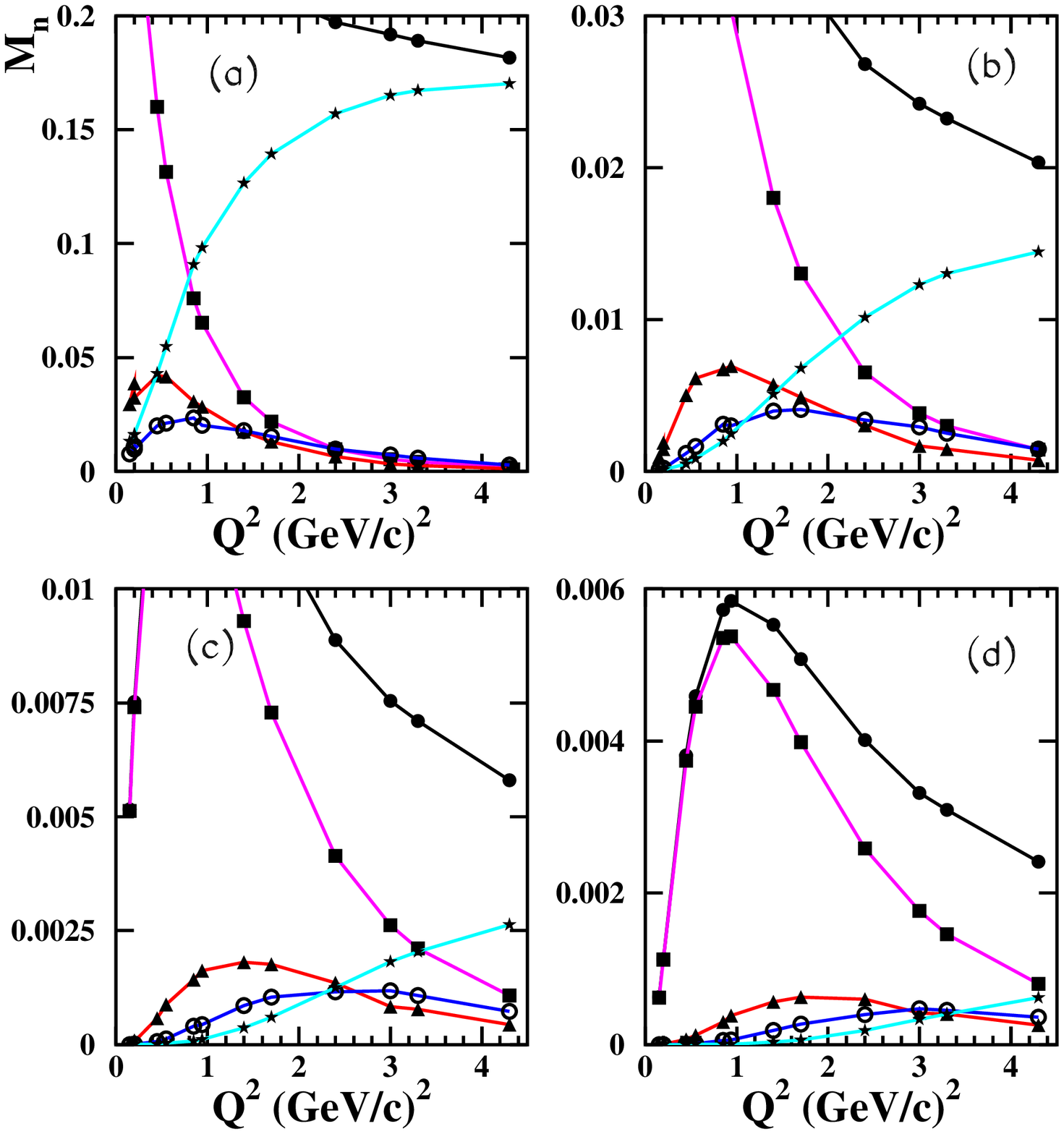}
\caption{Second (a), fourth (b), sixth (c) and eigth (d) Nachtmann
moments. Contributions due to the elastic peak (squares), the regions
1.2 $< W^2 <$ 1.9 GeV$^2$ (triangles),
1.9 $< W^2 <$ 2.5 GeV$^2$ (open circles), and
$W^2 >$ 4 GeV$^2$ (stars) are separately shown, in combination with
the total moment (solid circles), as a function of the momentum transfer.
Curves connect the various data, and are to guide the eye only.}
\end{center}
\end{figure}

\begin{figure}
\begin{center}
\epsfxsize=3.25in
\epsfysize=2.75in
\epsffile{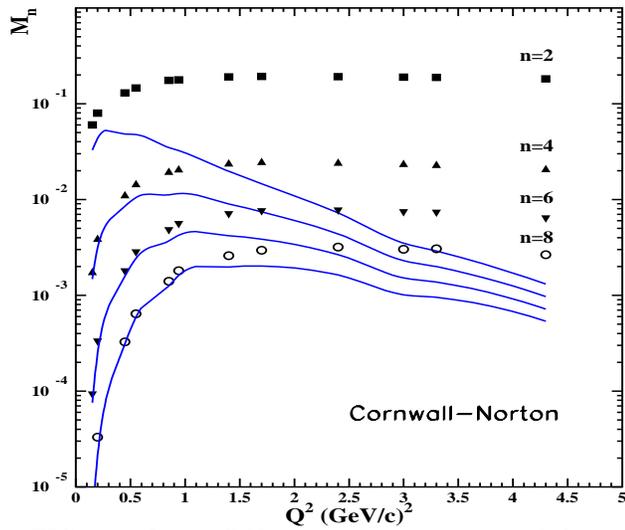}
\caption{Cornwall-Norton moments extracted from the world's electron-proton
scattering data, for $n$ = 2, 4, 6, and 8, {\bf without the elastic contribution
included}. The solid lines connect the calculated contributions from the
$N-\Delta$ transition region (1.2 $< W^2 <$ 1.9 GeV$^2$) at various $Q^2$.}
\end{center}
\end{figure}

\end{document}